\documentclass{rnaastex}
\pdfoutput=1

\newcommand{\oiii}{[O~{\footnotesize III}]}
\newcommand{\nii}{[N~{\footnotesize II}]}
\newcommand{\sii}{[S~{\footnotesize II}]}

\begin{document}

\title{The Planetary Nebula Luminosity Function (PNLF): Contamination from Supernova Remnants}

\author{Brian D. Davis}
\affiliation{Pennsylvania State University}

\author{Robin Ciardullo}
\affiliation{Pennsylvania State University}

\author{John J. Feldmeier}
\affiliation{Youngstown State University}

\author{George H. Jacoby}
\affiliation{Lowell Observatory}


\section{Introduction}

	Since the 1980's, \oiii\ $\lambda$5007 planetary nebula luminosity functions (PNLFs) have served as reliable standard candles for determining the distances of galaxies out to $\sim 20$ Mpc. The bright end of these PNLFs exhibit sharp exponential cutoffs, which are used in determining these distances. To this day, the method remains reliable and robust. A summary of the history of the PNLF is given in \cite{jacoby1992}.
	
	\cite{ferrarese2000} suggested that there may be a larger systematic error associated with the method than previously estimated for distances $> 10$ Mpc, claiming that at these distances, intracluster planetary nebulae (PNe) can lead to an underestimate of the bright-end cutoff magnitude by up to 0.2 mag. A rebuttal by \cite{ciardullo2002} argued that the apparent offset in distances may arise from a systematic difference between the internal extinction of nearby spiral galaxies and that within distant elliptical systems, and that the effect is also present in distances based on surface brightness fluctuations. Recent results from \cite{kreckel2017} have presented a third alternative: that the PNLFs of spiral galaxies at large distances are contaminated by supernova remnants (SNRs). Here we test this third hypothesis.

	A characteristic relationship exists between the \oiii\ $\lambda$5007 flux, H$\alpha$ flux, and \oiii\ $\lambda$5007 absolute magnitude of a population of PNe. \cite{herrmann2008} found that the vast majority of the PNe lie within a cone defined by $\log 4 > \log R > -0.37M_{5007} - 1.16,$ where $R = I$(\oiii\ $\lambda 5007$) / $I$(H$ \alpha$ + \nii) is the ratio of the \oiii\ $\lambda$5007 flux to the H$\alpha$ flux (whose bandpass also includes the \nii\ lines). Any spatially unresolved SNRs with flux ratios falling within this cone could be mistaken for PNe. At fixed seeing, this will occur more frequently at large distances. Thus our test: would the known SNRs in nearby spiral galaxies be classified as PNe if unresolved, and would these alter the PNLF bright-end cutoff magnitudes for these galaxies?

\section{Observations}
	
	We measure the narrowband fluxes of PNe and SNRs in M31 and M33. We choose these two particular galaxies for their proximity, and for the fairly complete PN and SNR surveys that have previously been performed in them. The PNe that we observe in M31 and M33 come from \cite{merrett2006} and \cite{ciardullo2004}, respectively. The SNR survey used for the M31 analysis is from \cite{lee2014}. In this survey, objects with roughly circular shapes and flux ratios $I$(\sii) / $I$(H$\alpha$) $> 0.4$ are classified as SNRs. For M33, our sample comes from the compilation of \cite{vucetic2015}, which also uses classification criteria based on the ratio of sulfur to hydrogen. Foreground extinctions for M31 and M33 are $A_V = 0.170$ and $A_V = 0.114$, respectively \citep{schlafly2011}.
	
	All photometric data for this study comes from the Lowell Local Group Survey, which uses the 4-meter telescope at Kitt Peak National Observatory. We use the B-band, H$\alpha$, and \oiii\ $\lambda$5007 images acquired for each galaxy, and measure PN and SNR fluxes using simple aperture photometry in the same manner as performed in \cite{ciardullo2002}.
	
\section{Results}
	
	In M31, we measure 446 PNe and 25 SNRs. In this sample, none of the SNRs fall within the PN cone. This means that none of the SNRs would be confused with PNe, even at distances $> 10$ Mpc.  The PNLF is therefore unaffected by contamination.
	
\begin{figure}[h!]
\begin{center}
\includegraphics[scale=0.9]{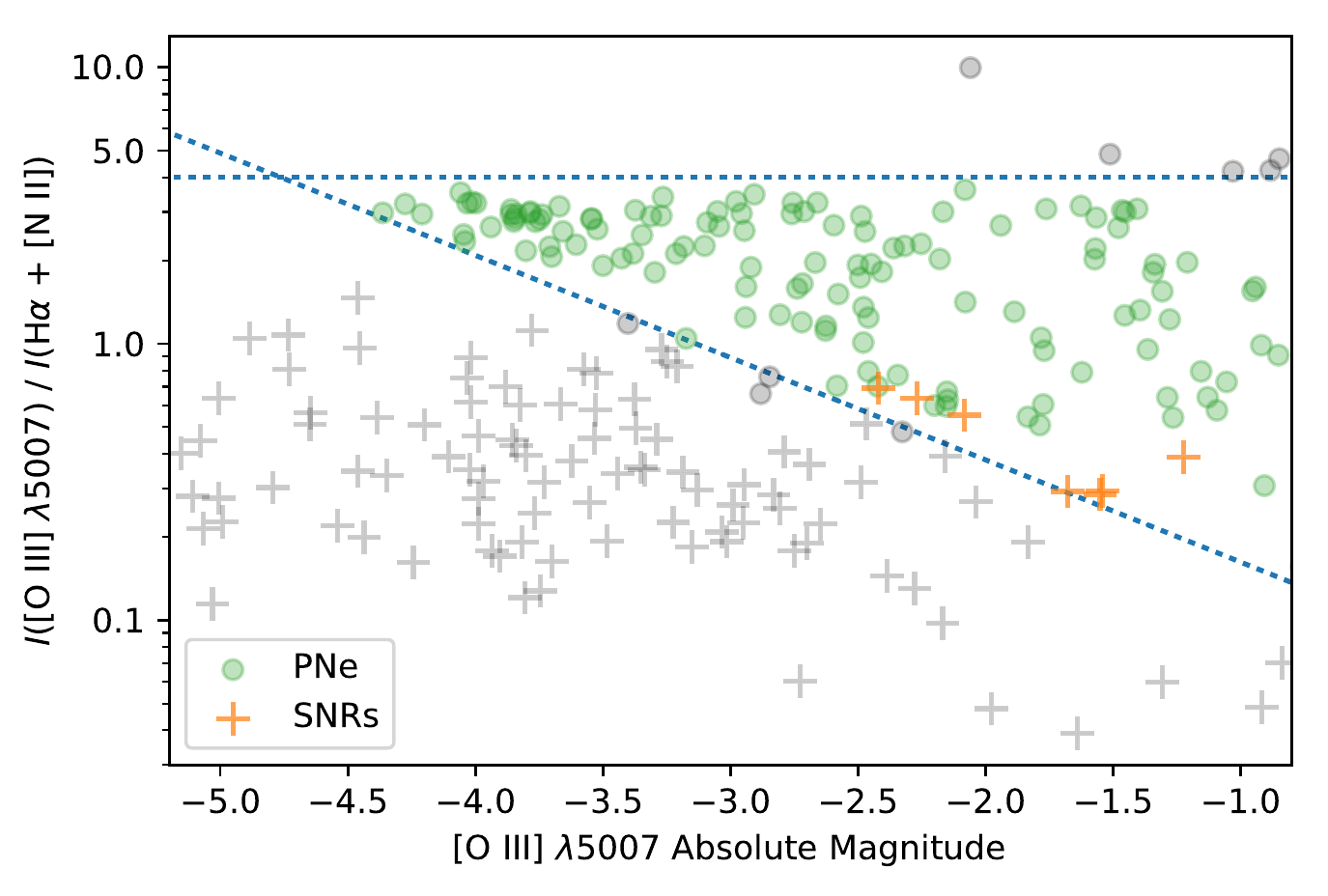}
\caption{PNe and SNRs in M33, represented by green circles and orange crosses, respectively. The dotted blue lines give the PN cone defined by \cite{herrmann2008}. Objects not within the standard PN cone are shown in gray. The vertical axis is the flux ratio of \oiii\ $\lambda$5007 to H$\alpha + $\nii. The PN measurements are from \cite{ciardullo2004}.}
\label{fig:m33_cone}
\end{center}
\end{figure}

	Our findings are only slightly different in M33. Figure \ref{fig:m33_cone} plots the flux ratios and \oiii\ $\lambda$5007 absolute magnitudes of 152 PNe and 112 SNRs in M33. Only 7 of the SNRs fall within the PN cone. These have $R<1$, and thus are not confused with the brightest PNe, which are 1 to 2 magnitudes brighter. In fact, no SNR (even outside of the cone) has an excitation indicative of a bright PN ($R \gtrsim 2$), and only four have $R > 1$. The bright end of the PNLF is unaffected by SNR contamination.
	
	In the case of M31 and M33, contamination of the PNLF by SNRs does not change the shape or location of the PNLF's bright-end cutoff. Because bright SNRs reside outside of the PN cone, PN/SNR confusion does not appear to be an important factor for galaxies at $>10$ Mpc, where compact SNRs are morphologically indistinguishable from PNe. It is possible that the result from \cite{kreckel2017} may have been somewhat of an anomaly.



\begin{thebibliography}{}

\bibitem[Ciardullo et al.(2002)]{ciardullo2002} Ciardullo, R., Feldmeier, J.~J., Jacoby, G.~H., et al.\ 2002, \apj, 577, 31 
\bibitem[Ciardullo et al.(2004)]{ciardullo2004} Ciardullo, R., Durrell, P.~R., Laychak, M.~B., et al.\ 2004, \apj, 614, 167 
\bibitem[Ferrarese et al.(2000)]{ferrarese2000} Ferrarese, L., Mould, J.~R., Kennicutt, R.~C., Jr., et al.\ 2000, \apj, 529, 745 
\bibitem[Herrmann et al.(2008)]{herrmann2008} Herrmann, K.~A., Ciardullo, R., Feldmeier, J.~J., \& Vinciguerra, M.\ 2008, \apj, 683, 630-643 
\bibitem[Jacoby et al.(1992)]{jacoby1992} Jacoby, G.~H., Branch, D., Ciardullo, R., et al.\ 1992, \pasp, 104, 599 
\bibitem[Kreckel et al.(2017)]{kreckel2017} Kreckel, K., Groves, B., Bigiel, F., et al.\ 2017, \apj, 834, 174 
\bibitem[Lee \& Lee(2014)]{lee2014} Lee, J.~H., \& Lee, M.~G.\ 2014, \apj, 786, 130 
\bibitem[Merrett et al.(2006)]{merrett2006} Merrett, H.~R., Merrifield, M.~R., Douglas, N.~G., et al.\ 2006, \mnras, 369, 120
\bibitem[Schlafly \& Finkbeiner(2011)]{schlafly2011} Schlafly, E.~F., \& Finkbeiner, D.~P.\ 2011, \apj, 737, 103 
\bibitem[Vu{\v c}eti{\'c} et al.(2015)]{vucetic2015} Vu{\v c}eti{\'c}, M.~M., Arbutina, B., \& Uro{\v s}evi{\'c}, D.\ 2015, \mnras, 446, 943 

\end{thebibliography}
\end{document}